\documentclass[sigconf, authorversion]{acmart}
\AtBeginDocument{%
  }

\copyrightyear{2026}
\acmYear{2026}
\setcopyright{cc}
\setcctype{by}
\acmConference[SAC '26]{The 41st ACM/SIGAPP Symposium on Applied Computing}{March 23--27, 2026}{Thessaloniki, Greece}
\acmBooktitle{The 41st ACM/SIGAPP Symposium on Applied Computing (SAC '26), March 23--27, 2026, Thessaloniki, Greece}
\acmDOI{10.1145/3748522.3779910}
\acmISBN{979-8-4007-2294-3/2026/03}

\usepackage{longtable}
\usepackage{graphicx}
\usepackage{listings}
\usepackage{makecell}
\usepackage{subcaption}
\usepackage{balance}

\def\Heap{{\mathcal H}}
\def\Ctx{{\mathcal E}}
\def\Par{{\tt par}}
\def\AT{{A}}
\def\Dirty{{\mathcal D}}
\def\True{{\it true}}
\def\False{{\it false}}
\def\If{{\tt if}}
\def\Then{{\tt then}}
\def\Else{{\tt else}}

\newcommand{\Ignore}[1]{}

\newtheorem{lemma}{Lemma}
\newtheorem{theorem}{Theorem}

\usepackage{listings}
\usepackage{xcolor}
\begin{document}

\title{The Essence of Entity Component System}
\renewcommand{\shorttitle}{The Essence of ECS}


\author{Anisha Tasnim} 
\affiliation{ \institution{University of Wisconsin-Milwaukee} \city{Milwaukee} \state{Wisconsin} \country{USA} }
\email{tasnim@uwm.edu} 
\author{Tian Zhao} \authornotemark[1] 
\affiliation{ \institution{University of Wisconsin-Milwaukee} \city{Milwaukee} \state{Wisconsin} \country{USA} }
\email{tzhao@uwm.edu} 

\renewcommand{\shortauthors}{A. Tasnim and T. Zhao}

\begin{abstract}
Modern game engines increasingly adopt the Entity Component System (ECS) paradigm as a data-oriented alternative to traditional object-oriented architecture. While ECS promotes modularity and performance through the separation of data and behavior, its practical efficiency depends heavily on the underlying data layout. Despite widespread adoption in frameworks, such as Unity DOTS, Bevy, and Flecs, the semantics of the archetype ECS remain informal and implementation-dependent, limiting rigorous reasoning about determinism, system scheduling, and structural mutations. 

This work formalizes and experimentally evaluates the archetype ECS. The formal model captures entity creation, component composition, system execution, and archetype migration as compositional state transitions, establishing the core invariants of archetype organization. Using a {\em Tower Defense} simulation, we compare the archetype ECS with alternative designs under identical conditions. 
Results show that the archetype ECS achieves higher frame rate and better frame stability than alternative designs, due to improved cache efficiency and consistent entity access. 
By uniting formal semantics with empirical validation, this study shows that the archetype ECS outperforms traditional architectures and provides a solid foundation for reasoning about correctness and parallelism. 
\end{abstract}



\maketitle

\section{Introduction}
Modern game engines must efficiently manage thousands of simultaneously active objects, such as towers, enemies, bullets, and visual effects along with maintaining smooth, real-time performance. In a {\em Tower Defense} game, for example, hundreds of towers track moving enemies, bullets fly through the air, and particle systems update simultaneously. Each frame must process every active entity's position, health, and state within milliseconds to maintain 60 frames per second (FPS) rendering speed. The way these entities are represented and accessed in memory critically affects performance.

Historically, most game engines have used Object-Oriented Programming (OOP), where each object (e.g., Tower, Enemy, Bullet) encapsulates its data and methods within hierarchical class structures. While intuitive, this design becomes inefficient as the number of entities increases. Each object instance occupies a scattered region of memory, leading to poor cache utilization when iterating over thousands of objects during game-play. For example, updating all enemy positions requires dereferencing each object individually, a process that repeatedly jumps across memory, resulting in cache misses and reduced throughput~\cite{Fedoseev2020,Hall2014}. As game complexity and hardware parallelism increase, these unpredictable access patterns become a fundamental bottleneck.

To address these limitations, the Entity Component System (ECS) architecture emerged, representing a shift toward Data-Oriented Design (DOD). In ECS, entities are simple identifiers, components are plain data structures (e.g., Position, Velocity, Health), and systems define the logic that operates over sets of components. For instance, in {\em Tower Defense}, a movement system updates all entities that have both {\tt Position} and {\tt Velocity} components, and a collision system processes \texttt{Health} components. This approach decouples data from behavior, allowing systems to operate over contiguous blocks of homogeneous data and enabling efficient parallelization~\cite{Harkonen2019,Romeo2016}.

However, the performance of ECS frameworks is deeply tied to their internal data layout. Early ECS implementations used an Array-of-Structs (AoS) layout, where each entity's components are grouped together in memory. This still leads to inefficient cache utilization during system-level iteration, such as systems only need one or two components from each entity. Struct-of-Arrays (SoA) layout improves this by storing each component type in its own contiguous array, enabling vectorized operations across entities and more predictable access patterns~\cite{Compton2022,bevy}.

Modern archetype-based ECS framework uses the SoA principle by grouping entities with identical component sets into archetypes, dense columnar tables where each column represents a component type and each row represents an entity. This design minimizes pointer chasing, enables  constant-time component access, and leverages CPU cache lines efficiently. Frameworks such as Unity DOTS, Bevy, and Flecs adopt this design to achieve significant performance improvements in large-scale simulations~\cite{bevy,flecs}. Moreover, recent work has extended ECS principles to high-performance domains. Madrona~\cite{Shacklett2023} used GPU to accelerate ECS for reinforcement learning environments, while Vico~\cite{HATLEDAL2021102243} used ECS for co-simulation across distributed systems, which demonstrated ECS's generality as a concurrent computation model.

Despite these advances, the semantics of archetype ECS frameworks remain informal, often described operationally without formal grounding. This lack of formalism makes it difficult to reason about determinism, system scheduling, and structural mutations. As ECS designs increasingly influence simulation engines and parallel runtime systems, establishing a formal semantics for archetype ECS becomes crucial. 

In this research, we make the following contributions:

\begin{enumerate}
    \item An operational semantics is defined to capture the core mechanisms of archetype ECS, such as entity creation, component association, system execution, and archetype migration in a compositional and deterministic manner. This semantics establishes the basic invariants of archetype-based data organization, showing that ECS execution can be described as a series of stable state transitions.

    \item A type system is defined for archetype ECS to ensure that a well-typed ECS program will not have unsafe access to archetype storage during runtime execution.
    
    \item An archetype SoA framework is implemented in Scala and its performance is compared with that of an object-oriented and an AoS implementation using a {\em Tower Defense} simulation~\cite{tower-defense-wiki}. The results demonstrate that the archetype SoA achieves higher frame stability.
\end{enumerate}

In the rest of the paper, we discuss related works in Section~\ref{sec:related} and present a motivational example in Section~\ref{sec:example}. In Section~\ref{sec:semantics}, we give an operational semantics to model the evaluation of ECS programs. Section~\ref{sec:type} defines a type system for ECS programs to prevent runtime errors due to unsafe access to archetype storage. Section~\ref{sec:implementation} describes an implementation of the {\em Tower Defense} simulation in ECS. Section~\ref{sec:performance} compares the performance of {\em Tower Defense} implemented in OOP, AoS, and archetype SoA, which shows that SoA has better game-play performance than the other two designs. 

\section{Related Work}
\label{sec:related}

Early game engines predominantly relied on object-oriented designs, where data and behavior were encapsulated within rigid inheritance hierarchies. These architectures encouraged modular encapsulation but scaled poorly as game logic became increasingly complex. The deep class hierarchies and runtime polymorphism typical of object-oriented designs produced non-deterministic memory access and poor cache utilization, leading to inefficiencies in highly parallel workloads~\cite{Fedoseev2020,Hall2014}. As hardware concurrency increased, these limitations prompted researchers to explore data-oriented designs that treat computation as transformations on structured memory layouts rather than class abstractions.

ECS emerged as a response to scalability challenges. ECS decomposes game logic into three dimensions: entities (identifiers), components, and systems (behavioral processors). This decomposition follows the principles of DOD, prioritizing predictable data access patterns and cache locality over inheritance-based flexibility~\cite{Harkonen2019,Romeo2016}. Unlike traditional object-oriented entities, ECS entities have no intrinsic behavior. In ECS, functionality arises through the combination of components and the systems that operate on them. This architectural separation enables composition over inheritance, enables runtime reconfiguration, and simplifies parallelization across systems. 

While ECS successfully decouples data and logic, the efficiency of its implementation depends primarily on the internal data layout used to store components. Early AoS models retained per-entity memory organization, where all components belonging to a single entity were stored contiguously. Although this design simplified access, it resulted in fragmented iteration patterns during system updates. In contrast, in SoA layouts, each component type is stored in a contiguous array. This improves spatial and temporal locality and enables Single Instruction Multiple Data (SIMD) vectorization, which accelerates iteration across large datasets~\cite{Compton2022}.

Modern archetype ECS architectures build upon the SoA model by grouping entities with identical component sets into archetypes. Each archetype is stored as a columnar table, where columns represent component types and rows represent entity instances. This organization eliminates pointer chasing and supports constant-time component lookup through column indexing~\cite{Compton2022}. Studies have demonstrated that this design achieves significant gains in update throughput and cache coherence compared to sparse-set ECS implementations~\cite{COX2025}. 

Several widely adopted frameworks embody these principles. Unity's DOTS architecture introduced chunk-based archetype storage with Burst-compiled jobs to mitigate cache inefficiency and main-thread contention~\cite{tichy2023}. Bevy ECS, implemented in Rust, integrates archetype tables with compile-time type safety through the Rust ownership model, achieving memory-safe, cache-optimized traversal~\cite{bevy}. Similarly, Flecs~\cite{flecs,flecs-github}, a C-based archetype ECS, organizes entities into dense tables and defers world modifications through command queues, ensuring thread-safe yet deterministic execution in parallel environments. These frameworks illustrate how archetype ECS designs maximize spatial locality and enable predictable and high-throughput updates suitable for large-scale simulations such as Tower Defense~\cite{tower-defense-github}.

The evolution of ECS has also expanded beyond game engines into broader high-performance and concurrent simulation contexts. Vico~\cite{HATLEDAL2021102243} applied ECS to distributed co-simulation, showing improved scalability through composition and fine-grained task isolation. Madrona~\cite{Shacklett2023} demonstrated a GPU-accelerated ECS runtime capable of executing large batched reinforcement learning environments as unified ``megakernels", achieving orders-of-magnitude speedups over CPU baselines. Cox {\em et al.}~\cite{COX2025} benchmarked archetype and sparse-set ECS designs using John Conway's {\em Game of Life}~\cite{Conway1970}, showing that archetype implementations nearly doubled iteration throughput at high entity counts, while sparse sets excelled in dynamic environments due to lighter update overhead. Fedoseev {\em et al.}~\cite{Fedoseev2020} compared Unity's object-oriented and DOD-based ECS prototypes, observing improved frame stability and reduced CPU load in the DOD version. These findings suggest that ECS designs and archetype-based SoA layouts in particular can achieve superior frame consistency and computational throughput in simulation-heavy workloads.

Despite extensive engineering refinement, the formal semantics of archetype ECS remain underdeveloped. Most existing frameworks describe their execution behavior operationally, relying on implementation heuristics rather than rigorous formalism. As ECS systems increasingly influence simulation engines, AI frameworks, and parallel schedulers, a formal model is necessary to reason about determinism, component migration, and synchronization invariants. Empirical comparisons reinforce the efficiency of archetype-based ECS systems. The absence of such semantics limits theoretical analysis and compiler-level optimization, motivating a structured, language-theoretic definition of ECS execution.

The most closely related work is Core ECS~\cite{Redmond2025}, which is a formalism that captures the semantics of ECS system focusing on concurrency and deterministic scheduling. It showed how safe schedule can be constructed and that schedule safety implies schedule determinism. In contrast, we model the archetype-based ECS architecture and provide an operational semantics along with type system that exposes precise read-write sets for every system. This enables us to reason statically about conflict freedom and system compatibility. While Core ECS focuses primarily on scheduling and deterministic concurrency, our approach provides a type-directed mechanism for detecting structural conflicts at compile time that is not expressible in Core ECS. Furthermore, our model directly reflects the concrete memory layout and archetype transitions used in real-world ECS engines. 

\section{Examples}
\label{sec:example}
This section introduces an example that illustrates the challenges addressed by our formalism. In particular, it demonstrates how ECS systems interact through shared archetypes, how structural mutations must be deferred to preserve iteration stability, and how read-write overlaps create the conflicts formalized in Section~\ref{sec:semantics}. 

\begin{figure}[ht!]
\centering
\includegraphics[width=1\linewidth]{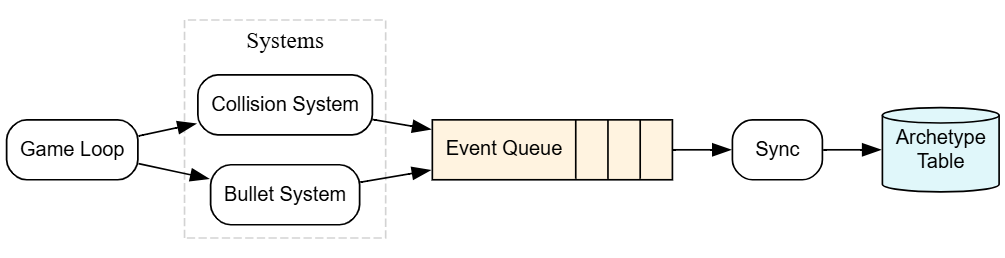}
\caption{Illustration of an ECS game loop, where queued events are synchronized to archetype table.}
\Description{Illustration of an ECS game loop, where queued events are synchronized to archetype table.}
\label{fig:ecs_flow}
\end{figure}

In {\em Tower Defense}, the efficiency of the game loop depends on how entities and their data are stored in memory. 
Figure~\ref{fig:ecs_flow} illustrates a part of the per-frame execution workflow of the ECS, which runs a {\tt BulletSystem} and a {\tt CollisionSystem}. The systems iterate over stable archetype tables, generate structural events, and defer these mutations until a subsequent synchronization step. The figure shows three main properties of archetype ECS:
(1) stable iteration sets, ensuring that systems observe a consistent snapshot of entity data;
(2) event staging, which defers all structural modifications until after system execution;
(3) all deferred mutations synchronized in an ordered batch to reestablish archetype consistency.
These mechanisms illustrate the interaction between entity iteration and structural mutation that our semantics captures through conflict modeling and tracking of archetypes dirtied by deferred mutations.

Each ECS system operates independently over entities with matching component sets. For example, \texttt{BulletSystem} in Listing~\ref{lst:bulletSystem} takes all entities with \texttt{Bullet} component. When a bullet exits the game boundary, it must be removed but directly modifying the world during iteration can corrupt archetype tables. To maintain structural integrity, such changes are issued as queued events and processed in a later synchronization (\texttt{world.sync()}) phase. This decoupling between event generation and structural mutation is essential for maintaining stable SoA memory layout and preventing mid-iteration inconsistencies.

\begin{lstlisting}[language=Scala, caption={Bullet system}, label={lst:bulletSystem}]

class BulletSystem(mapWidth=200f, mapHeight=200f) extends Systems {
  override val queryComponents = Set(classOf[Bullet])

  override def update(world: World, dt: Float): Unit = {
    val view = world.queryRows(queryComponents)

    view.foreach { r =>
      val b = r.table.getAt[Bullet](r.row, classOf[Bullet])

      b.x += b.dx * b.speed * dt
      b.y += b.dy * b.speed * dt
      b.ttl -= dt

      // Bounds & lifetime
      val outOfBounds =
        b.x < 0f || b.y < 0f || b.x > mapWidth || b.y > mapHeight

      if (b.ttl <= 0f || outOfBounds)
        world.enqueueEvent(Destroy(r.entity))
    }
  }
}
\end{lstlisting}

{\tt BulletSystem} and {\tt CollisionSystem}
illustrate how our semantics handles system interactions and detects conflicts. The bullet archetype contains entities with the components {\tt Bullet} and {\tt Position} and the enemy archetype contains entities with the components {\tt Position}, {\tt Health}, and {\tt Speed}. 
{\tt BulletSystem} iterates over the bullet archetype, reading each bullet's position and time-to-live, updating its state, and staging {\tt Destroy(bullet)} events when a bullet expires or leaves the map bounds. 
{\tt CollisionSystem} reads the enemy archetype to identify alive enemies, and reads the bullet archetype to test proximity between each bullet and its targeted enemy. When a hit is detected, it writes to the enemy archetype by decrementing enemy health and stages a {\tt Destroy(bullet)} event. 

\begin{figure}[ht!]
\centering
\includegraphics[width=0.8\linewidth]{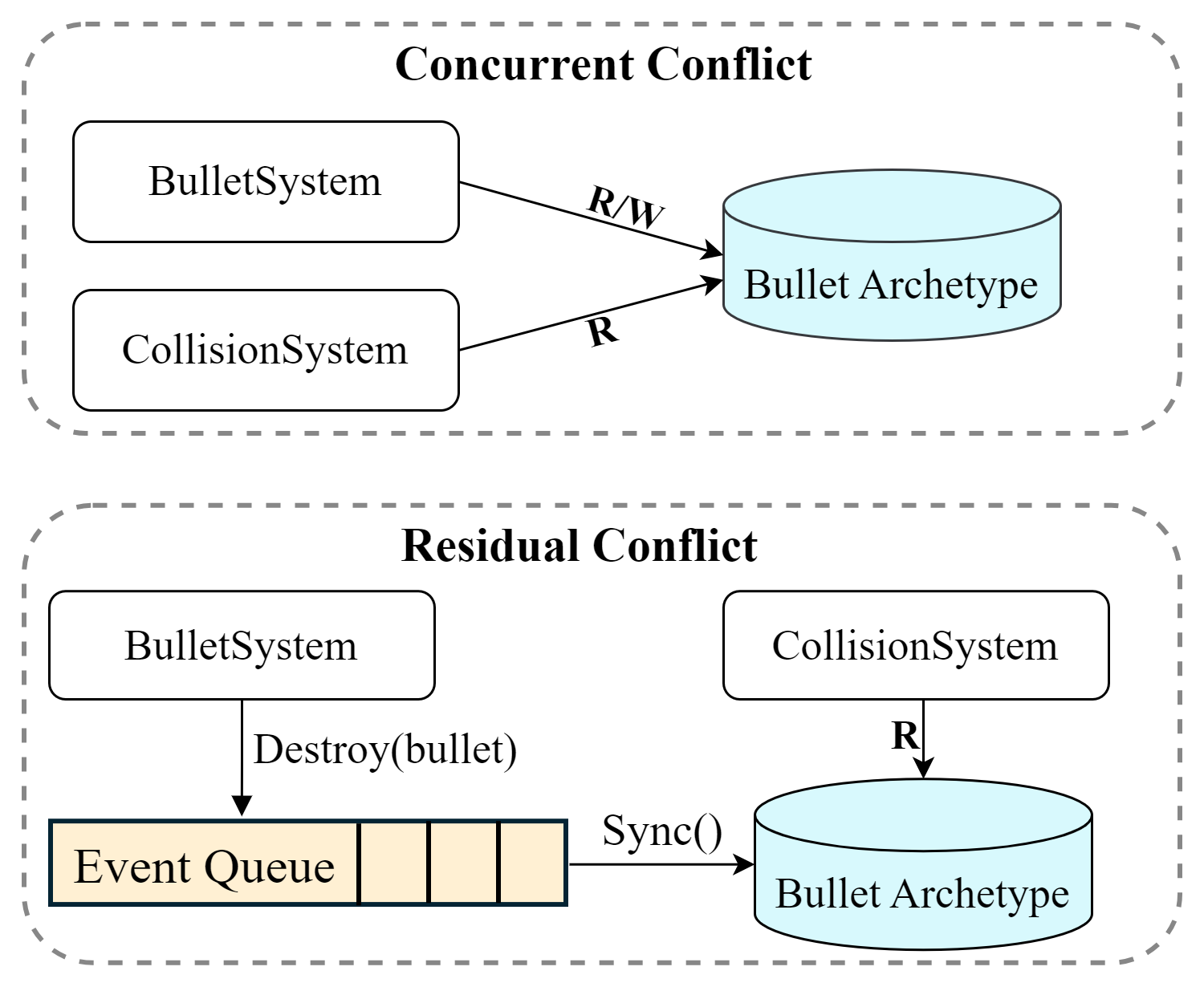}
\caption{Illustration of conflicts during the execution of {\tt BulletSystem} and {\tt CollisionSystem}.}
\Description{Illustration of conflicts during the execution of {\tt BulletSystem} and {\tt CollisionSystem}.}
\label{fig:conflict-parallel}
\end{figure}

As illustrated in Figure~\ref{fig:conflict-parallel}, though the two systems do not iterate the same entities, their access patterns over the bullet archetype produce conflicts. For example, {\tt BulletSystem} writes to the bullet archetype, while {\tt CollisionSystem} reads from it, creating a {\em concurrent conflict}. The systems also stage {\tt Destroy(bullet)} events, which leave the bullet archetype in a temporary dirty state until a synchronization step commits the structural updates, creating a {\em residual conflict}. To prevent concurrent conflict, {\tt BulletSystem} and {\tt CollisionSystem} cannot run in parallel. To prevent residual conflict, {\tt Destroy(bullet)} events must be synchronized before running systems that access the bullet archetype. 

\section{Semantics}
\label{sec:semantics}

In this section, we model the behavior of ECS programs via an operational semantics, which is a formal model describing how each ``step" in the system modifies stored data or interacts with entities and components. The semantics model terms, systems, and game state. By specifying precise semantic rules, we clarify how systems iterate over entities, change state, and maintain the ECS data reliably. Runtime errors include {\em conflicts} caused by unsafe access to archetypes in systems. A type system is defined for terms, systems, and game state so that a well-typed ECS program will not have conflicts and other runtime errors.

\subsection{Syntax}
Our ECS model adopts the syntax in Figure~\ref{fig:ecs} to formally describe entities, components, and systems. Primitive types $P$ represent the basic data types available in the ECS, such as Int, Float, and Boolean. 

Entities are represented by unique integers. Component types $C$ define data objects, each encapsulating multiple fields of primitive types $P_i$. Components consist of primitive values $v_i$ corresponding to their declared types. An archetype is a set of component types.

A system $(A,\lambda x.t)$ includes an archetype $\AT$ and a function that processes all entities of $\AT$. In $\lambda x.t$, $x$ is the variable for entities and $t$ is the operation performed on the entities and their components. Although a system can have multiple archetypes, this syntax only have one for simplicity. The systems can be sequential, data parallel, or task parallel. A sequential system will run in a single thread. A data parallel system can run in multiple threads by splitting the entities of the system among the threads. The task parallel system represents two or more systems running in parallel.

ECS libraries like Flecs~\cite{flecs} use query mechanism to retrieve archetypes that satisfy some query conditions.
We do not model query mechanism since the queries of task-parallel systems may return the same archetypes at runtime and safe treatment of queries complicates the formalism.

\begin{figure}[ht!]
\[\begin{array}{cclcll}
 P & \in & Primitive &=& Int \mid Float \mid \ldots \\[1mm]
 C & \in & CType  &=& \{P_1,\dots,P_n\}       & \mbox{Component type} \\[1mm]
 \AT & \in & AType  &=& \{C_1,\ldots,C_n\} & \mbox{Archetype} \\[1mm]
 e & \in & Entity &=& n                       & \mbox{Entity ID} \\[1mm]
 c & \in & Component &=& \{v_1,\dots,v_n\}    & \mbox{Component value} \\[1mm]
 s & \in & System &=& (\AT, \lambda x.t)    & \mbox{Sequential system} \\[1mm]
   &     &        &\mid& \Par~(\AT, \lambda x.t) & \mbox{Data parallel system} \\[1mm]
   &     &        &\mid& [s_1, s_2]  & \mbox{Task paralllel system}
\end{array}\]
\caption{The syntax of entities, components, and systems.}
\Description{A table showing the syntax of entities, components, archetypes, and systems in the ECS model.}
\label{fig:ecs}
\end{figure}

Term $t$ in Figure~\ref{fig:term-syntax} defines computational expressions within our ECS model. The terms include values, variables, conditionals, sequence, read and write entity components, and events.  
For simplicity, we omit computations such as function calls, local variables, and assignments.
Event $evt$ is the entity's life-cycle operations. An entity can be created or destroyed; a component can be added to or removed from an entity. 
The events may be staged ($lz~evt$) or immediate ($im~evt$). 
The immediate events can only be safely evaluated in a single-threaded system. Thus, most events are staged, which are placed in the event queue until they can be safely processed.

\begin{figure}[ht]
\[\begin{array}{cclcll}
t &\in& Term  &=& v           & \mbox{Value} \\[1mm]
  &   &       &\mid& x        & \mbox{Variable} \\[1mm]
  &   &       &\mid& \If~t~\Then~t_1~\Else~t_2 & \mbox{Conditional} \\[1mm]
  &   &       &\mid& t;~t'    & \mbox{Sequence} \\[1mm]
  &   &       &\mid& t.C      & \mbox{Read component} \\[1mm]
  &   &       &\mid& t.C := t'& \mbox{Update component} \\[1mm]
  &   &       &\mid& lz~evt   & \mbox{Staged event} \\[1mm]
  &   &       &\mid& im~evt      & \mbox{Immediate event} \\[2mm]
v &\in& Value &=& n           & \mbox{Primitive value} \\[1mm]
  &   &       &\mid& e        & \mbox{Entity} \\[1mm]
  &   &       &\mid& c        & \mbox{Component value}\\[2mm]  
evt&\in& Event&=& create~(C,c) & \mbox{Create entity} \\[1mm]
  &   &       &\mid& destroy~x& \mbox{Destroy entity} \\[1mm]
  &   &       &\mid& add~(x,C,c)&\mbox{Add component} \\[1mm]
  &   &       &\mid& remove~(x,C)&\mbox{Remove component}  
\end{array}\]
\caption{The syntax for terms, values, and events.}
\Description{}
\label{fig:term-syntax}
\end{figure}

\subsection{Archetype and Storage}
In an archetype ECS model, an entity is grouped with others sharing the same archetype, allowing for efficient processing.
Archetype stores similar entities in table-like storage, where each row is dedicated to an entity and component values are stored in columns. For convenience, we use $\Heap$ to represent the heap storage for archetypes, where $\Heap[\AT]$ returns the entities of an archetype $\AT$ and $\Heap[e][C]$ returns the component $C$ of entity $e$.

\[\begin{array}{cclcl}
 \Heap & \in & Heap & \stackrel{def}{=} & AType \rightarrow \{Entity\} \\[1mm]
       &     &      &\cup& Entity \rightarrow (Ctype \rightarrow Component) \\[2mm] 
\end{array}\]

\subsection{Game State and Frames}
The game state is defined as a tuple of heap, frame, and event queue, $(\Heap, fr, q)$. A frame contains the scheduled systems or synchronization steps that execute the staged events in the event queue.

\[\begin{array}{cclcll}
 w &\in& State &=& (\Heap, ~fr, ~q) & \mbox{Game state} \\[2mm]
 fr&\in& Frame &=& [sync] & \mbox{sync before next frame} \\[1mm]
   &   &       &\mid&  s::fr  & \mbox{run a system}~s \\[1mm] 
   &   &       &\mid&  sync::fr & \mbox{execute staged events} \\[2mm]
 q &\in& Queue  &=& [evt_1,\ldots,evt_n] & \mbox{Event queue} 
\end{array}\]

\subsection{Conflicts}
ECS supports data parallelism (a system runs in parallel threads where each thread operates on a separate set of entities) and task parallelism (multiple systems run in parallel). Since a system reads and/or writes the entities of archetypes or modify the archetypes (events), ECS must ensure that there is no conflicts between the threads that can lead to concurrency errors. 

The events cannot be executed in parallel since they modify archetype tables, which are not thread safe. Thus, events are often staged in a queue, which can be applied safely in a single thread between the run of the systems. 
If a system does not have immediate events, it is safe to run it in parallel threads since each thread operates on a separate partition of the archetype table and the staged events do not affect the archetype tables until they are applied.  

In practice, not all events can be staged. For example, in the {\tt CollisionSystem}, a bullet that collided with an enemy should be destroyed immediately so that the same bullet cannot hit multiple enemies. 
In this case, it is not safe run {\tt CollisionSystem} in parallel since race condition will occur due to concurrent modification to bullet archetype table.
Two or more systems can run in parallel if they do not have read/write access to the same archetype tables.

The staged events must be applied before running a system that depends on the archetypes modified by the events. For example, if a bullet entity is destroyed in {\tt CollisionSystem} but the render system runs before the destruction event is applied, then the destroyed bullet will be rendered.

In summary, there are two types of conflicts.
\paragraph{Concurrent Conflict} 
This occurs when the systems running in parallel threads have concurrent access to the same archetype tables (read/write the same entities or add/remove/modify entities). 
To avoid this, systems that have conflicting access to the same archetype tables cannot run in parallel. Also, a system with immediate events cannot run in parallel.

\paragraph{Residual Conflict} 
This occurs when a system uses an archetype that was dirtied by a previous system via staged events. 
Before executing a system, we must ensure that no archetype required by the system is dirtied. 
A synchronization barrier can also be placed before a system that accesses previously dirtied archetypes, ensuring that all staged events are fully applied and the archetype becomes clean again.

\subsection{Event Evaluation}

\begin{figure}[!ht]
\[\begin{array}{cr}
\dfrac{
    \begin{array}{c}
        e \mbox{ is a fresh integer} \quad
        \Heap_1 = \Heap[e \mapsto \{C \mapsto c\}] \\[1mm] 
        \AT = \{ C\} \quad
        \Heap_2 = \Heap_1[\AT \mapsto \Heap[\AT] \cup \{e\}]  
    \end{array}
    }{
        \Heap, create(C,c) \rightarrow \Heap_2, ()
    }
&
\mbox{\sc E-Cre}
\\[6mm]
\dfrac{
 \begin{array}{c}
   e \in \Heap[\AT] \quad
   \Heap' = \Heap[\AT \mapsto \Heap[\AT] - \{e\}]
  \end{array}
}{
  \Heap, destroy~ e \rightarrow \Heap', ()
}
&
\mbox{\sc E-Des} 
\\[6mm]
\dfrac{
  \begin{array}{c}
    e \in \Heap(\AT) \quad 
    \Heap_1 = \Heap[e \mapsto \Heap[e][C \mapsto c]] \\[1mm] 
    \Heap_2 = \Heap_1[\AT \mapsto \Heap_1[\AT] - \{e\}] \\[1mm]
    \AT_1 = \AT \cup \{C\} \quad
    \Heap_3 = \Heap_2[\AT_1\mapsto \Heap_2[\AT_1] \cup \{e\}] 
  \end{array}
}{
  \Heap, add(e, C, c)\rightarrow \Heap_3, ()
}
&
\mbox{\sc E-Add}
\\[6mm]
\dfrac{
 \begin{array}{c}
   e \in \Heap(\AT) \quad  
   \Heap_1 = \Heap[e \mapsto \Heap[e][C \mapsto \bot]] \\[1mm] 
   \Heap_2 = \Heap_1[\AT \mapsto \Heap_1[\AT] - \{e\}] \\[1mm]
   \AT_1 = \AT - \{C\} \quad
   \Heap_3 = \Heap_2[\AT_1\mapsto \Heap_2[\AT_1] \cup \{e\}] 
 \end{array}
}{
 \Heap, remove(e, C) \rightarrow \Heap_3, ()
}
&
\mbox{\sc E-Rem}
\end{array}\]
\caption{The rules for event reduction.}
\Description{}
\label{fig:event}
\end{figure}

Figure~\ref{fig:event} shows the evaluation rules for events, which include entity creation, entity destruction, and adding component to or removing component from an entity. $\Heap, evt \rightarrow \Heap', v$ represents the evaluation of $evt$ that changes the storage from $\Heap$ to $\Heap'$. 

A new entity is created with a component type and its initial value. When an entity is destroyed, it is removed from the archetype table where it was stored. Adding or removing a component from an entity changes the archetype of the entity. Thus, this entity is first removed from the old archetype table and then added to the new table. The component value of the entity is also updated.  

\subsection{Term Evaluation}

\begin{figure}[!ht]
\[\begin{array}{cr} 
\Heap, ~\If ~\True ~\Then ~t ~\Else ~t', ~q \rightarrow \Heap, ~t, ~q 
&
\mbox{\sc R-If-True}
\\[4mm]
\Heap, ~\If ~\False ~\Then ~t ~\Else ~t', ~q \rightarrow \Heap, ~t', ~q 
&
\mbox{\sc R-If-False}
\\[4mm]
\Heap, ~();t, ~q \rightarrow \Heap, ~t, ~q
&
\mbox{\sc R-Seq}
\\[4mm]
\dfrac{
 e \in \Heap[\AT] \quad C \in \AT 
}{
 \Heap, ~e.C, ~q \rightarrow \Heap, ~\Heap[e][C], ~q
}
&
\mbox{\sc R-Read-Comp}
\\[6mm]
\dfrac{
 \begin{array}{c}
 e \in \Heap[\AT] \quad C \in \AT \quad \\
 \Heap' =\Heap[e\mapsto \Heap[e][C\mapsto v]]
 \end{array}
}{
 \Heap, ~e.C:=v, ~q \rightarrow \Heap', ~(), ~q
}
&
\mbox{\sc R-Write-Comp}
\\[6mm]
\Heap, ~lz~evt, ~q \rightarrow \Heap, ~(), ~evt::q
&
\mbox{\sc R-Staged-Event}
\\[4mm]
\dfrac{
 \Heap, ~evt \rightarrow \Heap', v
}{
 \Heap, ~im~evt, ~q \rightarrow \Heap', ~v, ~q
}
&
\mbox{\sc R-Immedi-Event}
\\[6mm]
\dfrac{
 \Heap,~t,~q \rightarrow \Heap', ~t',~q' 
}{
 \Heap,~\Ctx[t],~q \rightarrow \Heap',~\Ctx[t'],~q'
}
&
\mbox{\sc R-Context} \\[4mm]
\end{array}\]
\[\begin{array}{rcl}
\Ctx[\cdot] &=& \cdot ~\mid~ \Ctx[\cdot];t ~\mid~ \If~\Ctx[\cdot]~\Then~t~\Else~t' \\[2mm] 
&\mid& \Ctx[\cdot].C ~\mid~ \Ctx[\cdot].C = t ~\mid~ e.C = \Ctx[\cdot]
\end{array}
\]
\caption{The rules for term reduction.}
\Description{}
\label{fig:term}
\end{figure}

Figure~\ref{fig:term} shows the small-step evaluation rules for terms, where  $\Heap, t, q \rightarrow \Heap', t', q'$ reduces term $t$ to $t'$ with possibly new storage and queue.
Most rules are standard including reading/writing components of an entity.
A staged event $lz~evt$ ($lz$ is short for lazy) is added to the event queue. An immediate event $im~evt$ is evaluated to a value resulting a new archetype storage.

\subsection{System Evaluation}

\begin{figure}[!ht]
\[\begin{array}{cr}
\dfrac{
\begin{array}{c}
 es = [e_1,\ldots,e_n] \\[1mm]
 \forall i \in [1..n].~\Heap_{i-1}, ~[e_i/x]t, ~[] \rightarrow^* \Heap_i, v_i, ~q_i
\end{array}
}{
  \Heap_0, \lambda x.t ~@~ es \rightarrow \Heap_n, ~q_1+\ldots+q_n
}
&
\mbox{\sc S-Iter} 
\\[6mm]
\dfrac{
  es = \Heap[\AT] \quad
  \Heap, \lambda x.t ~@~ es \rightarrow \Heap', ~q 
}{
  \Heap,~(\AT, \lambda x.t) \rightarrow \Heap', ~q
}
&
\mbox{\sc S-Seq}
\\[6mm]
\dfrac{
\begin{array}{c}
  es = es_1 \cup es_2 = \Heap[\AT] \quad es_1\cap es_2=\emptyset \\[1mm]
  i \in \{1,2\}.~ 
  \Heap,~\lambda x.t~@~es_i \rightarrow \Heap_i,~ q_i 
\end{array}
}{
  \Heap, ~\Par~(\AT, \lambda x.t) \rightarrow 
  merge(\Heap, \Heap_1, \Heap_2), ~q_1 + q_2
}
&
\mbox{\sc S-Data-P}
\\[6mm]
\dfrac{
\begin{array}{c}
  \forall i\in\{1,2\}.~\Heap, ~s_i \rightarrow \Heap_i,~q_i \\[1mm]
  query(s_1) \cap query(s_2) = \emptyset
\end{array}
}{
  \Heap, ~[s_1, s_2] \rightarrow
  merge(\Heap, \Heap_1, \Heap_2), ~q_1 + q_2
}
&
\mbox{\sc S-Task-P}
\\[8mm]
\dfrac{
 \Heap, ~s \rightarrow \Heap', ~q' \quad
 no\_conflict(\Heap, s, q)
}{
 \Heap, ~s::fr,~ q \rightarrow \Heap',~ fr, ~q + q'
}
&
\mbox{\sc F-System}
\\[6mm]
\dfrac{
\begin{array}{cr}
  \forall evt_i \in q=[evt_1,\ldots,evt_n]. ~\Heap_{i-1}, ~evt_i \rightarrow \Heap_i, ()
\end{array}
}{
  \Heap_0, ~sync :: fr, ~q \rightarrow \Heap_n, ~fr, ~nil
}
&
\mbox{\sc F-Sync}
\end{array}\]
\caption{The rules for system and frame evaluation.}
\Description{}
\label{fig:system}
\end{figure}

\begin{figure}[!ht]
\[\begin{array}{c}
\begin{array}{c}
query(A,\lambda x.t) = \{A\} \quad
query(\Par~s) = query(s) \\[2mm]
query([s_1,s_2]) = query(s_1) \cup query(s_2)
\end{array}
\\[8mm]
\dfrac{
\begin{array}{c}
 \forall \AT.~\Heap'[\AT] = \Heap[\AT] = \Heap_1[\AT] = \Heap_2[\AT] \\[1mm]
 \forall e,i. \mbox{ if } \Heap_i[e] \neq \Heap[e] \mbox{ then }
 \Heap'[e] = \Heap_i[e]  \mbox{ else } \Heap'[e] = \Heap[e]
\end{array}
}{
 merge(\Heap, \Heap_1, \Heap_2) = \Heap'
}
\\[6mm]
\begin{array}{c}
\dfrac{
 \AT \notin atype(\Heap, q) 
}{
 no\_conflict(\Heap, ~(\AT,\lambda x.t), ~q) 
}
\\[6mm]
\dfrac{
 no\_conflict(\Heap, s, q)
}{
 no\_conflict(\Heap, \Par~s, q) 
}
\quad
\dfrac{
 \forall i \in\{1,2\}.~no\_conflict(\Heap, s_i, q)
}{
 no\_conflict(\Heap, [s_1,s_2], q) 
}
\end{array}
\\[12mm]
\begin{array}{lcl}
  atype(\Heap, create(C, c)) &=& \{\{C\}\} \\[2mm]
  atype(\Heap, destroy~e) &=& \{\AT\} \mbox{ where } e \in \Heap(\AT) \\[2mm]
  atype(\Heap, add(e,C,c)) &=& \{\AT, \AT \cup \{C\}\} \mbox{ where } e \in \Heap(\AT) \\[2mm]
  atype(\Heap, delete(e,C)) &=& \{\AT, \AT - \{C\}\} \mbox{ where } e \in \Heap(\AT) \\[2mm]
  atype(\Heap, q) &=& \bigcup_{evt\in q} atype(\Heap, evt)
\end{array}
\end{array}\]
\caption{Auxiliary functions.}
\Description{}
\label{fig:auxiliary}
\end{figure}

Figure~\ref{fig:system} shows the evaluation rules for systems, which can be sequential, data parallel, or system parallel.
A system $(\AT, \lambda x.t)$ is evaluated by calling $\lambda x.t$ with each entity of the archetype $\AT$. 
Rule ({\sc S-Iter}) describes how $\lambda x.t$ iterates over entity list $es$, denoted by $(\lambda x.t)@es$. The notation $\rightarrow^*$ means transitive closure of $\rightarrow$. 

\paragraph{Sequential system} A system $s = (\AT,\lambda x.t)$ runs over the entities $es$ of $\AT$. In the sequential mode, it simply iterates over each entity in $es$, applying $t$, and possibly staging events. 

\paragraph{Data parallel system} If a system needs to process a large number of entities, it can process partitions of the entities in separate threads. For simplicity, Rule ({\sc S-Data-P}) only shows two partitions. Each thread returns a new archetype heap $\Heap_i$ and queue $q_i$. The queues are combined. The heaps are merged, where the archetype tables must not change and only the component values of entities may be updated. This rule precludes immediate events in parallel threads.

\paragraph{Task parallel system} 
A list of systems can run in parallel threads. The restriction is similar to that of the data parallel systems except that task parallel systems have different archetypes. For simplicity, Rule ({\sc S-Task-P}) only shows two systems.

\subsection{Frame Reduction}

A game state is a triple of archetype heap $\Heap$, a frame $fr$, and an event queue $q$. A frame is a list of systems and sync operations. The transition of a state is the execution of the system or sync on top of the frame, which is defined in Figure~\ref{fig:system}. 
The predicate $no\_conflict(\Heap, s, q)$ ensures that the entities of the archetype accessed by $s$ is not dirtied by events in $q$.  

The sync operation applies all queued events in order and then clears the event queue. This design is chosen for simplicity. In practice, an ECS library can automatically execute events if they are in conflict with the next system in the frame.
When a frame completes, the game loop can restart with the same frame or dynamically reconfigure the frame with new systems. 

\section{Type System}
\label{sec:type}

Our ECS model has a simple set of types $\tau$, which include primitive types, unit type, component types, and archetypes, which are sets of component types.

\[\begin{array}{ccrl}
   \tau &=& P  & \mbox{Primitive type} \\
        &\mid& () & \mbox{Unit type} \\
        &\mid& C & \mbox{Component type -- set of primitive types} \\
        &\mid& \AT & \mbox{Archetype -- set of component types}
\end{array}\]

\begin{figure}[!ht]
\[\begin{array}{cr}
\Gamma \vdash x:\Gamma(x)~ \& ~\emptyset
&
\mbox{\sc T-Var}
\\[4mm]
\dfrac{
  \Gamma \vdash ~t_1:\tau_1 ~\&~ W_1 \quad \Gamma \vdash t_2 :\tau_2 ~\& ~W_2
}{
  \Gamma \vdash (t_1;t_2) : \tau_2  ~\& ~W_1 \cup W_2
}
&
\mbox{\sc T-Seq}
\\[6mm]
\dfrac{
  \Gamma \vdash t: Bool \quad 
  \Gamma \vdash t_1: \tau ~\&~ W_1 \quad 
  \Gamma \vdash t_2: \tau ~\&~ W_2
}
{
\If ~t ~ \Then ~t_1 ~\Else ~t_2 : \tau  ~\&~ W_1\cup W_2 
}
&
\mbox{\sc T-If}
\\[6mm]
\dfrac{
  C \in \Gamma(x) 
}{
  \Gamma \vdash ~x.C : C ~\&~ \emptyset
}
\quad \quad
\dfrac{
  C \in \Gamma(x) \quad 
  \Gamma \vdash t: C ~\& ~W
}{
  \Gamma \vdash x.C = t : C ~\& ~W
}
&
\mbox{\sc T-Comp}
\\[6mm]
\dfrac{
  \Gamma \vdash evt : () ~\& ~W
}{
  \Gamma \vdash lz~evt : () ~\& ~W
}
\quad \quad
\dfrac{
  \Gamma \vdash evt : ()~\& ~W
}{
  \Gamma \vdash im~evt : () ~\& ~\emptyset
}
&
\mbox{\sc T-Event}
\end{array}\]
\caption{Term typing rules.}
\Description{}
\label{fig:term-type}
\end{figure}

\begin{figure}[!ht]
\[\begin{array}{cr}
\Gamma \vdash create(C,c) : () ~\& ~\{ \{C\} \}
&
\mbox{\sc T-Create}
\\[4mm]
\dfrac{
 \Gamma \vdash t: \tau ~\&~ W
}{
 \Gamma \vdash destroy ~ t: () ~\& ~W \cup \{\tau\}
}
&
\mbox{\sc T-Destroy}
\\[6mm]
\dfrac{
 \Gamma \vdash t: \tau ~\&~ W \quad C \notin \tau \quad 
 \tau' = \tau \cup \{C\}
}{
 \Gamma \vdash add~(t, C, c) : () ~\& ~W \cup \{\tau,\tau'\}
}
&
\mbox{\sc T-Add}
\\[6mm]
\dfrac{
 \Gamma \vdash t: \tau ~\&~ W \quad C \in \tau \quad 
 \tau' = \tau - \{C\}
}{
 \Gamma \vdash remove~(t, C) : () ~\& ~W \cup \{\tau,\tau'\}
}
&
\mbox{\sc T-Remove}
\end{array}\]
\caption{Event typing rules.}
\Description{}
\label{fig:event-type}
\end{figure}

\begin{figure}[!ht]
\[\begin{array}{cr}
\dfrac{
 \begin{array}{c}
   x: \AT \vdash t: \tau ~\&~ W
   \quad t \mbox{ has no staged events}
 \end{array}
}{
 \vdash (\AT,~\lambda x.t) : (\{\AT\}, W)
} 
& 
\mbox{\sc T-Seq}
\\[6mm]
\dfrac{
  \vdash s: (R, W) \quad s \mbox{ has no immediate events}
}{
 \vdash \Par~s : (R, W)
} 
&
\mbox{\sc T-Data-P}
\\[6mm]
\dfrac{
 \begin{array}{c}
  s_1,~s_2 \mbox{ have no immediate events} \\
  \vdash s_1: (R_1, W_1) \quad \vdash s_2: (R_2, W_2) \quad
  R_1 \cap R_2 = \emptyset 
 \end{array}
}{
\vdash [s_1, s_2] : (R_1\cup R_2, W_1 \cup W_2)
}
&
\mbox{\sc T-Task-P}
\\[6mm]
\dfrac{
 \vdash fr: \emptyset  
}{
 \vdash (sync::fr) : \Dirty
}
&
\mbox{\sc T-Sync}
\\[6mm]
\dfrac{
   \vdash s: (R, W) \quad 
   R \cap \Dirty = \emptyset \quad
   \vdash fr: W \cup \Dirty
}{
   \vdash (s::fr) : \Dirty  
} 
&
\mbox{\sc T-System}
\end{array}\]
\caption{Frame and system typing rules.}
\Description{}
\label{fig:frame-type}
\end{figure}

We use a type and effect system to track the read and write operations to archetypes during system computation.
The type judgment for term has the form $\Gamma \vdash t: \tau ~\&~ W$, where given the variable environment $\Gamma$, $\tau$ is the type of the term $t$ and $W$ is the set of archetypes that may be written by staged events in $t$.
The typing rules in Figure~\ref{fig:term-type} and~\ref{fig:event-type} collect the archetypes of the staged events and put them in the write set $W$. The archetypes of the immediate events are not tracked since they are evaluated sequentially.

The type judgment for systems has the form $\vdash s: (R, W)$, which says that the system $s$ will access entities of archetypes in $R$ and produce staged events that can change archetypes in $W$. 
Rule ({\sc T-Seq}) checks the type of a sequential system, where the parameter $x$ has the type $\AT$ since the entities passed to $x$ are in the archetype $\AT$. Mixing immediate and staged events in the same system can be problematic since immediate events may change the heap structure that the staged events depend on. Thus, for simplicity, we require sequential system to have no staged events while data/task parallel systems to have no immediate events.

The type rules for frames in Figure~\ref{fig:frame-type} tracks the set of dirtied archetypes $\Dirty$. Rule ({\sc T-System}) ensures that the archetypes in $R$ are not in $\Dirty$ and combines the write set $W$ with $\Dirty$ in checking the rest of the frame. Rule ({\sc T-Sync}) clears $\Dirty$ for systems after a sync. 

\subsection{Properties}

\begin{figure}[ht!]
\[\begin{array}{cr}
\Gamma \vdash e : \Gamma(e) ~\&~ \emptyset \quad \quad
\dfrac{
 \Gamma \vdash t: \tau ~\&~ W \quad W \subseteq W'
}{
 \Gamma \vdash t: \tau ~\&~ W'
}

& \mbox{\sc T-Sub}
\\[6mm]
\dfrac{
 \forall e_i \in es. \quad e_i \in \Heap[\AT] \quad 
 e_i: \AT \vdash [e_i/x] t: \tau ~\&~ W
}{
 \vdash \Heap, \lambda x.t ~@~ es : (\{A\}, W)
}
& \mbox{\sc T-Iter}
\\[6mm]
\dfrac{
 \vdash c_1:C_1 \ldots \vdash c_n:C_n
}{
 \vdash \{C_1:c1,\ldots,C_n:c_n\} : \{C_1,\ldots,C_n\} 
}
& \mbox{\sc T-Entity} \\[6mm]
\dfrac{
 \forall \AT.~\forall e\in\Heap[\AT].~\vdash \Heap[e]: \AT
}{
 \vdash \Heap
}
& \mbox{\sc T-Heap} \\[4mm]
\dfrac{
  \vdash \Heap \quad
  \Dirty \supseteq atype(\Heap, q) \quad
  \vdash fr: \Dirty
}{
  \vdash \Heap, fr, q 
}
&
\mbox{\sc T-Frame}
\end{array}\]
\caption{Type rules for runtime values.}
\Description{Type rules for runtime values.}
\label{fig:runtime}
\end{figure}

In this section, we state the progress and type preservation lemmas to show that a well-typed ECS program will not get stuck. 
The proof, omitted here, uses the typing rules of runtime values in Figure~\ref{fig:runtime}. 
Rule ({\sc T-Frame}) says that a game state $\Heap, fr, q$ is well-typed if $\Heap$ is well-typed and the frame $fr$ is well-typed with respect to the archetypes dirtied by $q$.
Rules ({\sc T-Heap}) and ({\sc T-Entity}) check that every entity of every archetype is well-typed.

\begin{lemma}[Progress]\label{lemma:progress}
If $\vdash \Heap, fr, q$, then there exists $\Heap'$, $fr'\neq fr$, and $q'$ such that $\Heap, fr, q \rightarrow \Heap', fr', q'$.
\end{lemma}

The interesting part of the proof is that if $fr = s::fr'$, then for $s$ to reduce, it cannot access the archetypes dirtied in $q$. This can be shown from the typing rules for frame and system.  

\begin{lemma}[Preservation]\label{lemma:preservation}
If $\vdash \Heap, fr, q$ and $\Heap, fr, q \rightarrow \Heap', fr', q'$, then $\vdash \Heap', fr', q'$.
\end{lemma}

The interesting part of the proof is that if $fr = s::fr'$ and $\vdash s: (R,W)$, then $W$ includes the archetypes dirtied by staged events in $s$. Together with the typing rules for frame and system, we can show that the new game state is well typed. 

\begin{theorem}[Soundness]
If $\vdash \Heap, fr, q$, then there exists $\Heap'$ such that 
$\Heap, fr, q \rightarrow^* \Heap', [], nil$.
\end{theorem}
The progress and preservation lemmas ensure that a well-typed game state will evaluate its frame until it is empty. Since a frame always ends with a $sync$, the event queue will be cleared in the end. 

\section{Experimental Implementation}
\label{sec:implementation}

We implemented a {\em Tower Defense} simulation to compare four designs: OOP, Array-of-Structs (AoS) ECS, and single/multi-threaded archetype Struct-of-Arrays (SoA) ECS. Archetype SoA organizes entities into archetypes (sets of component types) and stores their data in column-major layout, enabling efficient bulk operations. All structural mutations, including component addition/removal and entity creation/destruction, are deferred and queued during system execution, and synchronized before executing the next system.

\subsection{Tower Defense Architecture}
{\em Tower Defense} models waves of enemies moving along a path while stationary towers automatically detect and attack them. The simulation advances in a deterministic main loop that runs at fixed time steps. Each frame executes a predefined sequence of systems and synchronization points. Systems are pure functions that operate over component data matched by a query. All game state resides in components, which are plain data classes without embedded behavior accessed via stable entity IDs. 

\subsection{Archetype Storage}
In archetype SoA, each archetype is defined as a unique set of component classes, mapped to an {\tt ArchetypeTable}, which is a columnar store that maintains all entities sharing that exact component sets. A global index maps each entity to its current table and row, enabling $O(1)$ component lookup. Structural mutations to archetypes may be deferred as queued events, which can be synchronized before the next system executes. During synchronization, affected entities are migrated between archetype tables using a swap-and-pop algorithm to maintain dense storage. Once all structural updates have been applied, the archetype-query cache is invalidated so that subsequent systems operate on up-to-date data.

\subsection{Core Systems}
The game is driven by a sequence of systems, each operating over component arrays selected by its query signature. A system declares the components it reads or writes, and at runtime the ECS engine identifies the matching archetype tables and executes the system’s update function directly over the relevant columns. 

{\em Tower Defense} includes position updates ({\tt MovementSystem}), combat logic ({\tt BulletSystem}, {\tt CollisionSystem}, {\tt ShootingSystem}), enemy spawning ({\tt EnemySpawning}), enemy health ({\tt HealthSystem}), collision effects ({\tt ParticleSystem}), a path following algorithm ({\tt PathFollowingSystem}) and visualization ({\tt RenderSystem}). Each system is implemented as a function that takes the current world state and delta time, and mutates only the components it explicitly queries. For example, {\tt MovementSystem} iterates over all entities with {\tt Position} and {\tt Velocity}, updating coordinates by applying the velocity scaled by delta time.

The execution order of systems is defined statically, which sequences system steps and synchronization points. This design enables fine-grained control over update dependencies and makes system behavior reproducible. Dynamic scheduling of systems is possible for more complex behavior though each new schedule must be verified at runtime for safety.

\subsection{Query Interface}
In archetype SoA, the systems access the archetype tables through a query mechanism. A system's query specifies the components it requires and at runtime the engine resolves this signature to the archetype tables that contain exactly those components.

To ensure performance and cache locality, the query mechanism looks up entities using archetype indices via hash maps and the system receives a map from component type to data, enabling in-place mutation. The query interface provides critical functionality to systems, enabling them to retrieve the number of matching entities, iterate efficiently over entity IDs alongside their component data. Query results are stable within a single frame execution, the interface ensures that iteration sets remain valid throughout the execution of each system.

\subsection{Parallelism}
SoA-PAR is our parallel version of the archetype SoA design. It includes data and task parallel execution over archetype-structured storage. 

\paragraph{Task Parallelism} 
SoA-PAR employs task parallelism at the system level to exploit coarse-grained concurrency across independent subsystems. Each system encapsulates a distinct unit of game logic operating over a subset of components. These systems declare their memory access patterns through an access descriptor that specifies which component types are read, written, or structurally modified during execution.

At runtime, {\tt FrameExecutorParallel} class analyzes each system's access sets and organizes them into waves of conflict-free execution. Systems whose read-write do not overlap are grouped into the same wave, and executed concurrently using the shared worker pool managed by the Scheduler. Systems that perform structural updates are placed in their own serialized waves to ensure a consistent view of archetype state. 

The scheduler executes waves sequentially. After running all systems in a wave in parallel, it immediately invokes {\tt sync()} to apply the staged events. Each wave boundary serves as a synchronization point for both execution and structural consistency, ensuring that subsequent waves observe a fully updated world state. This design provides deterministic ordering, eliminates residual conflicts across waves, and achieves scalable parallelism for system-level tasks.

\paragraph{Data Parallelism}
Within each system, data parallelism is implemented with a lightweight construct, {\tt ParFor}, which partitions an entity range into contiguous chunks distributed across worker threads. Each thread executes the same update function on its assigned subset, achieving SIMD-like parallelism. 

For instance, the {\tt MovementSystem} updates entity positions by applying a uniform computation over all entities with position and velocity. Using {\tt ParFor.parFor}, this loop is partitioned into chunks that execute concurrently on all available processor cores. The chunk size is dynamically chosen based on the number of hardware threads and a configurable granularity parameter, balancing load distribution and scheduling overhead. This model yields near-linear speedup for data-oriented systems dominated by arithmetic or memory-bound operations.

In summary, SoA-PAR supports two-level parallelism:
\begin{enumerate}
    \item Task-level parallelism between systems whose data dependencies permit concurrent execution.
    \item Data-level parallelism within each system across independent entity records.
\end{enumerate}
By nesting these two forms of concurrency, the runtime maximizes CPU utilization on multi-core architectures while preserving deterministic execution.

\section{Performance Comparison}
\label{sec:performance}
We compared OOP, AoS, archetype SoA, archetype SoA-PAR by running {\em Tower Defense} implemented in each design, where only SoA-PAR uses multiple threads. Although it is well established in the literature and in industrial practice that ECS outperforms OOP, our experiments quantify the performance gain obtained specifically from an implementation based on our ECS semantics. We use OOP as the baseline because OOP remains the common architecture in game development and interactive simulations, especially in commercial engines and educational materials. By comparing against OOP, we show how much of the performance improvement comes directly from our formal model.

We ran each design under the same game-play conditions (max entities 20000, max enemies 15000, enemy spawn interval 0.05s, turret firing interval 0.02s), targeting a fixed frame rate of 60 FPS. Performance data were collected through an integrated profiler. Experiments were conducted on a system equipped with Intel Core i7-12650H CPU, 32 GB DDR5 RAM, and NVIDIA GeForce RTX 4060 GPU. All ECS computation and system scheduling run exclusively on the CPU and the GPU is used only for rendering.
To ensure stability of the performance, each configuration was executed multiple times. Across these runs we observed minimal variance, and all runs produced consistent timing behavior. Because the differences were negligible, we report a representative run for each configuration.

We evaluated two foreground loop policies in {\tt LibGDX/LWJGL}: an uncapped run with {\tt foregroundFPS=0}, which performs no throttling and renders as fast as the hardware allows, and a capped run with {\tt foregroundFPS=60}, where the engine uses cooperative sleep/yield to target 60 FPS without syncing to the monitor. Each configuration was executed several times and the variance of the results were minimal. A full scalability study with varying entity count and different map size is deferred to future work.

\paragraph{Frame Rate}
Table~\ref{tab:avg-fps-comparison-par} represents the average FPS achieved among four designs under the two rendering settings. 
OOP exhibits the lowest performance in both settings and shows marked frame-time instability. The AoS design nearly doubles OOP throughput, while the archetype SoA yields an additional improvement. The parallel variant (SoA-PAR) performs the best overall, sustaining over 100 FPS with an unconstrained foreground rate and maintaining close to the 60 FPS target under capped conditions.

\begin{table}[ht]
\centering
\caption{Average FPS across architectures \& foreground FPS.}
\label{tab:avg-fps-comparison-par}
\resizebox{\columnwidth}{!}{
\begin{tabular}{lcc}
\hline
\textbf{Architecture} & \makecell{\textbf{Avg FPS}\\\{\em foreground FPS = 0\}} & \makecell{\textbf{Avg FPS}\\\{\em foreground FPS = 60\}} \\
\hline
OOP & 36.13 & 25.49 \\
AOS & 61.51 & 47.71 \\
Archetype SoA & 65.81 & 48.76 \\
Archetype SoA-PAR & 109.54 & 58.54 \\
\hline
\end{tabular}
}
\end{table}

\begin{figure*}[ht]
\centering
\begin{subfigure}[b]{0.65\linewidth}
    \centering
    \includegraphics[width=\linewidth]{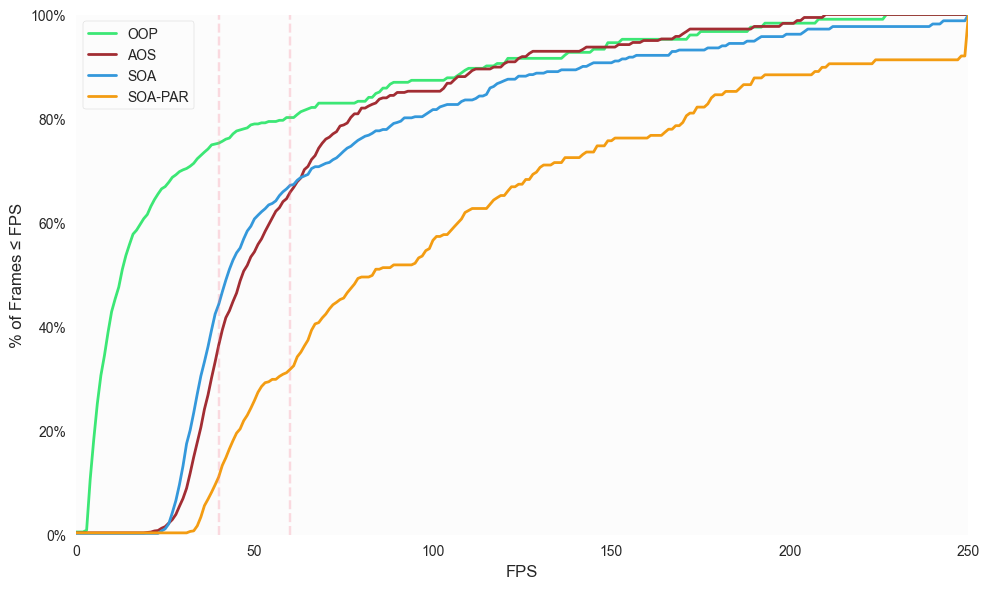}
    \caption{Cumulative FPS distribution ({\em foregroundFPS=0}).}
    \label{fig:cdf-fps-foregrund-0}
\end{subfigure}
\hfill
\begin{subfigure}[b]{0.65\linewidth}
    \centering
    \includegraphics[width=\linewidth]{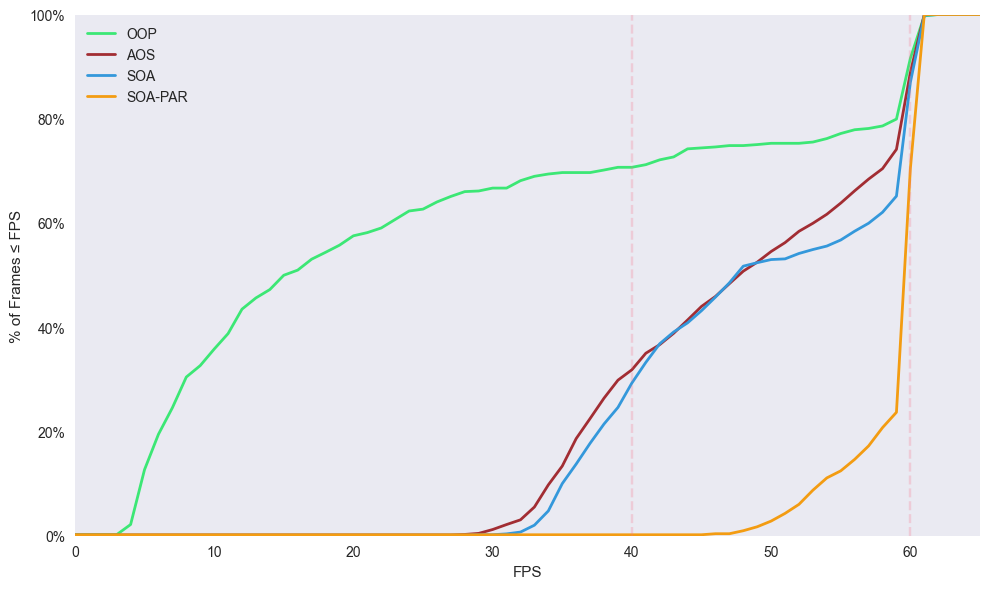}
    \caption{Cumulative FPS distribution ({\em foregroundFPS=60}).}
    \label{fig:cdf-fps-foregrund-60}
\end{subfigure}
\caption{Comparison of cumulative FPS distributions for OOP, AoS, SoA, SoA-PAR with or without capping FPS at 60.}
\Description{Comparison of cumulative FPS distributions for the OOP, AoS, archetype SoA, SoA-PAR with or without frame-rate capped at 60 FPS.}
\label{fig:cdf-combined}
\end{figure*}

The cumulative FPS distributions in Figure~\ref{fig:cdf-combined} illustrates the stability differences among architectures. Under the uncapped setting ({\em foregroundFPS=0}), shown in Figure~\ref{fig:cdf-fps-foregrund-0}, SoA-PAR consistently dominates the entire CDF curve, reflecting both higher throughput and smoother frame pacing. The standard SoA configuration ranks second, outperforming AoS and OOP due to improved cache locality and reduced memory misses.
In the capped setting ({\em foregroundFPS=60}), shown in Figure~\ref{fig:cdf-fps-foregrund-60}, the SoA-PAR curve rises steeply near the 60 FPS mark, indicating minimal variance and stable frame timing close to the target rate. By contrast, OOP exhibits a much flatter distribution, revealing substantial frame-to-frame fluctuations and degraded temporal stability.

\section{Conclusion}
In this paper, we presented a formal semantics and a type system for an archetype-based ECS framework. Our semantic model captures the essence of archetype ECS computation in that entities can be read and updated in parallel by stateless systems while mutations to archetype tables are delayed until they can be safely processed. 
We also evaluated the performance of four architectural designs, including OOP, AoS, archetype SoA, and its parallel extension (SoA-PAR) using a {\em Tower Defense} simulation. The results align with prior findings: the SoA architecture consistently outperforms OOP and AoS due to its contiguous memory layout and improved cache locality. The parallel SoA design further amplifies these benefits by exploiting multi-core hardware, achieving both higher throughput and stable frame pacing under load. 

As future work, we will conduct a scalability study with varying entity counts, game maps, and system complexity. 
We will also extend our formalism to model system query to retrieve entities using filters. While queries are flexible, they introduce challenges to static checking of task-parallel systems since the sets of queried archetypes are dynamic. Another interesting feature is entity relationships~\cite{flecs}. For example, {\tt Turret} entities can relate to {\tt Bullet} entities via a {\tt Fires} relationship, which enables more capable queries than component-based ones. However, cleaning up relationships after entity destruction is a challenge.  
Other design choices include automatic execution of staged events based on the archetypes of the next system and dynamic scheduling of parallel systems based on their dependencies and the availability of threads~\cite{Romeo2016}. 

The source code is available at \url{https://github.com/uwm-se/ecs}.

\begin{acks}
This work is partially supported by the Northwestern Mutual Data Science Institute (NMDSI) under grant number SS136.
\end{acks}

\balance

\bibliographystyle{ACM-Reference-Format}
\bibliography{ref}

\end{document}